\documentclass[a4paper,11pt,english,oneside]{article}

\usepackage[english]{babel}
\usepackage[latin1]{inputenc}
\usepackage[T1]{fontenc}
\usepackage{amsmath}
\usepackage{amsfonts}
\usepackage{amssymb}
\usepackage{verbatim}
\usepackage{rotating}
\usepackage{textcomp}
\usepackage{colortbl}
\usepackage{xcolor}
\usepackage{graphicx}
\usepackage{fancyhdr}
\usepackage{algorithm} 
\usepackage{algorithmic}
\usepackage{amsthm}
\usepackage{indentfirst}
\usepackage[usenames,dvipsnames]{pstricks}
\usepackage{epsfig}

\usepackage[letterpaper]{geometry}
\usepackage[all]{xy}

\newcommand{\FF}{\mathbb{F}}
\newcommand{\KK}{\mathbb{K}}
\newcommand{\ZZ}{\mathbb{Z}}
\newcommand{\QQ}{\mathbb{Q}}
\newcommand{\RR}{\mathbb{R}}
\newcommand{\CC}{\mathbb{C}}
\newcommand{\NN}{\mathbb{N}}

\newcommand{\A}{{\mathcal A}}
\newcommand{\B}{{\mathcal B}}

\newcommand{\NLI}{{\mathcal N}}

\newcommand{\nP}{{\mathfrak n}}

\newcommand{\w}{\mathrm{w}}
\newcommand{\N}{\mathrm{N}}

\newcommand{\op}{{\sf p}} 
\newcommand{\Bf}{B.f.\;}

\newcommand{\ef}{\underline{f}}
\newcommand{\eg}{\underline{g}}

\newcommand{\dist}{\mathrm{d}}

\newtheorem{theorem}{Theorem}[section]

\newtheorem{definition}[theorem]{Definition}

\newtheorem{proposition}[theorem]{Proposition}

\begin{document}
\date{}
\title{Yet another algorithm to compute the nonlinearity of a Boolean function}
\author{
E.~Bellini\\
{\footnotesize Department of Mathematics, University of Trento, Italy}\\
{\footnotesize eemanuele.bellini@gmail.com}
}
\maketitle
\begin{abstract}
We associate to each Boolean function a polynomial whose evaluations represents the distances from all possible Boolean affine functions. Both determining the coefficients of this polynomial from the truth table of the Boolean function and computing its evaluation vector requires a worst-case complexity of  $O(n2^n)$ integer operations. This way, with a different approach, we reach the same complexity of established algorithms, such as those based on the fast Walsh transform.
\end{abstract}

\begin{center}
\end{center}



\section{Introduction}
  \label{secIntro}
Any function from $(\FF_2)^n$ to $\FF_2$ is called a Boolean function. Boolean functions are important in symmetric cryptography, since they are used in the confusion layer of ciphers. An affine Boolean function does not provide an effective confusion. To overcome this, we need functions which are as far as possible from being an affine function. The effectiveness of these functions is measured by several parameters, one of these is called ``nonlinearity''.\\
In this paper, we provide a method to compute the nonlinearity of a Boolean function,
estimating its complexity compared to that of the classical methods which uses fast Walsh and fast Mobius transform.
\\
In Section \ref{secPrelOnBF} we recall the basic notions and theorems we need.
In Section \ref{secNLwithFPE} we associate to each Boolean function in $n$ variables a polynomial whose evaluations represent the distance from all possible affine functions, yielding an algorithm to compute the nonlinearity. We also provide a theorem to express the coefficients of this polynomials.
Finally, in Section \ref{secNLComplexity} we analyze the complexity of the proposed method, both experimentally and theoretically. 
In particular, we arrive at a worst-case complexity of  $O(n2^n)$ operations over the integers, that is, sums and doublings. This way, with a different approach, we reach the same complexity of established algorithms, such as those based on the fast Walsh transform.


\section{Preliminaries and Notation on Boolean functions}
  \label{secPrelOnBF}

In this chapter we summarize some definitions and known results from \cite{CGC-cd-book-carlet} and \cite{CGC-cd-book-macwilliamsI}, concerning \Bf and the classical techniques to determine their nonlinearity.\\

Let $\FF$ denote the field $\FF_2$. The set $\FF^n$ is the set of all binary vectors of length $n$, viewed as an $\FF$-vector space.\\
Let $v\in\FF^n$. The \emph{Hamming weight} $\w(v)$ of the vector $v$ is the number of its nonzero coordinates. For any two vectors $v_1,v_2\in\FF^n$, the \emph{Hamming distance} between $v_1$ and $v_2$, denoted by $\dist(v_1,v_2)$, is the number of coordinates in which the two vectors differ.\\
A \emph{Boolean function} (\Bf) is any function $f:\FF^n\rightarrow \FF$. The set of all \Bf's from $\FF^n$ to $\FF$ will be denoted by $\B_n$.\\
%
\indent
We assume implicitly to have ordered $\FF^n$, so that $\FF^n=\{\op_1,\ldots,\op_{2^n}\}$.\\ 
A \Bf $f$ can be specified by a \emph{truth table}, which gives the evaluation of $f$ at all $\op_i$'s.
%
We consider the evaluation map from $\B_n$ to $\FF^{2^n}$, associating to each \Bf $f$ the vector $\underline{f}=(f(\op_1)\ldots,f(\op_{2^n}))$, which is called the \emph{evaluation vector} of $f$.
Once the order on $\FF^n$ is chosen, i.e. the $\op_i$'s are fixed, it is clear that the evaluation vector of $f$ uniquely identifies $f$.\\
%
%
\indent
A \Bf $f\in\B_n$ can be expressed in a unique way as a polynomial in $\FF[X]=\FF[x_1,\ldots,x_n]$, as
$f=\sum_{v \in \FF^n}b_vX^v\,,$
where $X^v=x^{v_1}\cdots x^{v_n}$.
This representation is called the \emph{Algebraic Normal Form} (ANF).\\
%
%
%
Let $\A_n = \{\alpha \in \B_n \mid \alpha(X)=a_0 + \sum_{i=1}^na_ix_i, (a_0,\ldots,a_n) \in \FF^{n+1} \}$ denote the set of all \emph{affine} functions.
\\
%
%
%
\indent
In \cite{CGC-cry-art-carlet1999} a useful representation of \Bf's is introduced for characterizing several cryptographic criteria.\\
\Bf's can be represented as elements of $\KK[X]/\langle X^2-X \rangle$, where $\langle X^2-X \rangle$ is the ideal generated by the polynomials $x_1^2-x_1,\ldots,x_n^2-x_n$, and $\KK$ is $\ZZ$, $\QQ$, $\RR$, or $\CC$.
 Let $f$ be a function on $\FF^n$ taking values in a field $\KK$. We call the \emph{numerical normal form} (NNF) of $f$ the following expression of $f$ as a polynomial:
 $
 f(x_1,\ldots,x_n) = \sum_{u \in \FF^n}\lambda_u (\prod_{i=1}^{n}x_i^{u_i}) = \sum_{u \in \FF^n}\lambda_{u}X^u\,,
 $
 with $\lambda_{u} \in \KK$ and $u=(u_1,\ldots,u_n)$.\\
It can be proved 
that any \Bf $f$ admits a unique NNF, consideiring values in $\KK$.\\
%
\indent
From now on let $\KK = \QQ$. 
The truth table of $f$ can be recovered from its NNF by the formula $f(u)=\sum_{a\preceq u}\lambda_a,\forall u \in \FF^n\,,$
where $a\preceq u\iff \forall i \in \{1,\ldots,n\} \; a_i \le u_i$. Conversely, 
it is possible to derive an explicit formula for the coefficients of the NNF by means of the truth table of $f$.
\begin{proposition}\label{propNNFcoeff}
 Let $f$ be any integer-valued function on $\FF^n$. For every $u\in \FF^n$, the coefficient $\lambda_u$ of the monomial $X^u$ in the NNF of $f$ is:
 \begin{equation}\label{eqNNFCoeff}
  \lambda_u = (-1)^{\w(u)}\sum_{a\in \FF^n |a\preceq u}(-1)^{\w(a)}f(a)\,.
 \end{equation}
\end{proposition}



%
Let $f,g\in\B_n$. The distance $\dist(f,g)$ between $f$ and $g$ is the number of $v\in\FF^n$ such that $f(v)\neq g(v)$.
%
It is obvious that
$\dist(f,g)=\dist(\ef,\eg)=\w(\ef+\eg)\,.$
%
\begin{definition}
Let $f\in\B_n$. The \emph{nonlinearity} of $f$ is the minimum of the distances between $f$ and any affine function, i.e.
$\N(f)=\min_{\alpha\in\A_n}\dist(f,\alpha)\,.$
\end{definition}
%
%
%
\indent
Using a simple divide-and-conquer butterfly algorithm it is possible to compute the ANF from the truth-table of a \Bf, by performing $O(n2^{n})$ bit operations, and storing $O(2^n)$ bits. This algorithm is known as the \emph{fast M\"obius transform}. From the ANF it is possible to compute the nonlinearity of a \Bf by means of a similar algorithm known as the \emph{fast Walsh transform}, requiring $O(n2^n)$ integer sums and 
storing $O(2^n)$ integers.\\
\indent
Faster methods are known in particular cases, for example when the ANF is a sparse polynomial \cite{CGC-cry-phdthesis-calik2013}.

%
%



%


\section{Computing the nonlinearity using fast polynomial evaluation}
  \label{secNLwithFPE}
%

Let $A$ be the variable set $A=\{a_i\}_{0\leq i\leq n}$. We denote by $\mathfrak{g}_n\in\FF[A,X]$
the following polynomial:
$$\mathfrak{g}_n=a_0+\sum_{i=1}^n a_i x_i\,
\,.$$
Determining the nonlinearity of $f\in\B_n$ is the same as finding the minimum weight of the vectors in the set $\{\ef+\eg\mid g\in\A_n\}\subset\FF^{2^n}$.
We can consider the evaluation vector of the polynomial $\mathfrak{g}_n$ as follows (see \cite{CGC-cd-art-ilawcc07}):
$$\underline{\mathfrak{g_n}}=(\mathfrak{g}_n(A,\op_1),\ldots,\mathfrak{g}_n(A,\op_{2^n}))\in (\FF[A])^{2^n}\,.$$
%
%
From now on we present original results.\\
For each $0 \le i \le 2^n$, we define the following Boolean affine polynomials:
$$f_{i}^{(\FF)}(A)=\mathfrak{g}_n(A,\op_i)+f(\op_i)\,.$$
We also define
$$f_i^{(\ZZ)}(A) = \text{NNF}(f_{i}^{(\FF)}(A))\,.$$
%
\begin{definition}\label{defNLP}
 We call $\nP_f(A) = f_1^{(\ZZ)}(A)+\dots+f_n^{(\ZZ)}(A) \in \ZZ[A]$ the {\bf integer nonlinearity polynomial} (or simply the \emph{nonlinearity polynomial}) of the \Bf $f$.\\
 For any $t\in \NN$ we define the ideal $\NLI_f^t \subseteq \QQ[A]$ as follows:
 \begin{align}
  \NLI_f^t =
  \langle E[A] \bigcup \{ f_1^{(\ZZ)}+\dots+f_{2^n}^{(\ZZ)}-t \} \rangle =
  \langle E[A] \bigcup \{ \nP_f-t \} \rangle
 \end{align}
\end{definition}
 Notice that the integer evaluation vector $\underline{\nP_f}$ represents all the distances of $f$ from all possible affine functions in $n$ variables, and so the following theorem is straitforward.
\begin{theorem}
The variety of the ideal $\NLI_f^t$ is non-empty if and only if
the Boolean function $f$ has distance $t$ from an affine function. 
In particular, $\N(f) = t$, where $t$ is the minimum positive integer such that $\mathcal{V}(\NLI_f^t)\ne \emptyset$.
\end{theorem}
%
%
Thus, to compute the nonlinearity of $f$ we
have to find the minimum nonnegative integer $t$ in the set of the evaluations of $\nP_f$, that is, in $\{\nP_f(\bar{a}) \mid \bar{a} \in \{0,1\}^{n+1} \subset \ZZ^{n+1}\}$.\\
Now we claim a theorem to calculate the coefficients of the nonlinearity polynomial. Proof is omitted due to lack of space (see \cite{CGC-cry-art-BellSimSala14}).
\begin{theorem}\label{thmNPCoeffFormula}
  Let 
  $v=(v_0,v_1,\ldots,v_n)\in\FF^{n+1}$,
  $\tilde{v}=(v_1,\ldots,v_n)\in\FF^{n}$, 
  $A^v=a_0^{v_0} \cdots a_n^{v_n} \in \FF[A]$ 
  and 
  $c_v \in \ZZ$ be such that
  $\nP_f = \sum_{v\in\FF^{n+1}}c_vA^v$. 
  Then the coefficients of $\nP_f$ can be expressed as:
  \begin{align}  
  \label{eqCoeff0}
  c_{v} = \sum_{u\in \FF^{n}}f(u) = \w(\underline{f}) \,\,\text{ if } v = 0
  \\
  \label{eqCoeffv}
  c_v =
  (-2)^{\w(v)}
  \sum_{\substack{u\in \FF^n \\ \tilde{v}\preceq u}} 
  \left[
  f(u) - \frac{1}{2}
  \right]
  \,\,\text{ if } v \ne 0
  \end{align}
%
\end{theorem}

\section{Complexity considerations}
  \label{secNLComplexity}
%
We write the algorithm (Algorithm \ref{algNLPcoeff}) to calculate the nonlinearity polynomial in $O(n2^n)$ integer operations.\\
In Figure \ref{figNLPn3} Algorithm \ref{algNLPcoeff} is shown for $n=3$.
\begin{algorithm}[H]
{\footnotesize
\caption{Algorithm to calculate the nonlinearity polynomial  $\nP_f$ in $O(n2^n)$ integter operations.}
\label{algNLPcoeff}
  \begin{algorithmic}[1]
    \REQUIRE{The evaluation vector $\underline{f}$ of a \Bf $f(x_1,\ldots,x_n)$}
    \ENSURE{the vector $c=(c_1,\ldots,c_{2^{n+1}})$ of the coefficients of $\nP_f$} \\
    Calculation of the coefficients of the monomials \emph{not} containing $a_0$
    \STATE{$(c_1,\ldots,c_{2^n}) = \underline{f}$}
    \FOR{$i=0,\ldots,n-1$}
      \STATE{$b\leftarrow 0$}
      \REPEAT
        \FOR{$x = b,\ldots,b+2^i-1$}
        \STATE{$c_{x+1} \leftarrow c_{x+1} + c_{x+2^i+1}$}\label{stepSUM} 
        \IF{$x = b$}
          \STATE{$c_{x+2^i+1} \leftarrow 2^i -2c_{x+2^i+1}$}\label{stepG1} 
        \ELSE
          \STATE{$c_{x+2^i+1} \leftarrow -2c_{x+2^i+1}$}\label{stepG2} 
        \ENDIF
        \ENDFOR
        \STATE{$b \leftarrow b+2^{i+1}$} 
      \UNTIL{$b=2^n$}
    \ENDFOR \\
  Calculation of the coefficients of the monomials containing $a_0$
  \STATE{$c_{1+2^n} \leftarrow 2^n -2c_{1}$}
  \FOR{$i = 2,\ldots,2^n$}
    \STATE{$c_{i+2^n} \leftarrow-2c_{i}$}
  \ENDFOR
    \RETURN $c$
  \end{algorithmic}
} 
\end{algorithm}
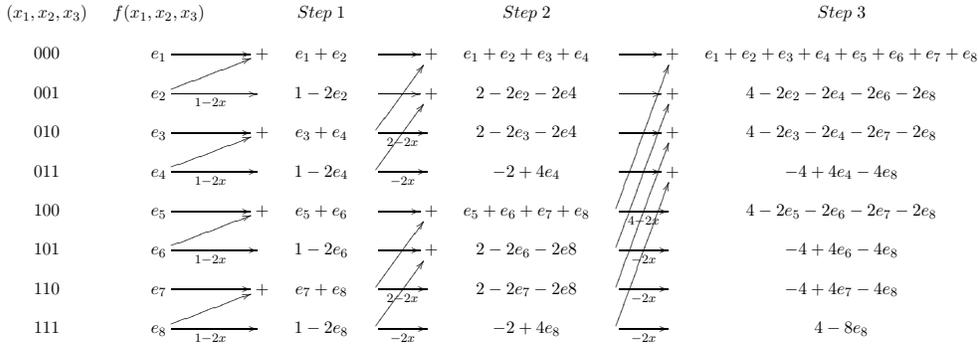
\begin{figure}[h]
\scalebox{.6}{
 \xymatrix@=10pt{
  (x_1,x_2,x_3) & f(x_1,x_2,x_3)           & &   & Step\;1     &                         & &   & Step\;2                 &                           & &   & Step\;3                                          & \\  
  000        & e_{1}\ar[rr]                & & + & e_{1}+e_{2} & \ar[rr]                 & & + & e_{1}+e_{2}+e_{3}+e_{4} & \ar[rr]                   & & + & e_{1}+e_{2}+e_{3}+e_{4}+e_{5}+e_{6}+e_{7}+e_{8}  & \\
  001        & e_{2}\ar[urr]\ar[rr]_{1-2x} & &   & 1-2e_{2}    & \ar[rr]                 & & + & 2-2e_{2}-2e{4}          & \ar[rr]                   & & + & 4-2e_{2}-2e_{4}-2e_{6}-2e_{8}                    & \\
  010        & e_{3}\ar[rr]                & & + & e_{3}+e_{4} & \ar[uurr]\ar[rr]_{2-2x} & &   & 2-2e_{3}-2e{4}          & \ar[rr]                   & & + & 4-2e_{3}-2e_{4}-2e_{7}-2e_{8}                    & \\
  011        & e_{4}\ar[urr]\ar[rr]_{1-2x} & &   & 1-2e_{4}    & \ar[uurr]\ar[rr]_{-2x}  & &   & -2+4e_{4}               & \ar[rr]                   & & + & -4+4e_{4}-4e_{8}                                 & \\
  100        & e_{5}\ar[rr]                & & + & e_{5}+e_{6} & \ar[rr]                 & & + & e_{5}+e_{6}+e_{7}+e_{8} & \ar[uuuurr]\ar[rr]_{4-2x} & &   & 4-2e_{5}-2e_{6}-2e_{7}-2e_{8}                    & \\
  101        & e_{6}\ar[urr]\ar[rr]_{1-2x} & &   & 1-2e_{6}    & \ar[rr]                 & & + & 2-2e_{6}-2e{8}          & \ar[uuuurr]\ar[rr]_{-2x}  & &   & -4+4e_{6}-4e_{8}                                 & \\
  110        & e_{7}\ar[rr]                & & + & e_{7}+e_{8} & \ar[uurr]\ar[rr]_{2-2x} & &   & 2-2e_{7}-2e{8}          & \ar[uuuurr]\ar[rr]_{-2x}  & &   & -4+4e_{7}-4e_{8}                                 & \\
  111        & e_{8}\ar[urr]\ar[rr]_{1-2x} & &   & 1-2e_{8}    & \ar[uurr]\ar[rr]_{-2x}  & &   & -2+4e_{8}               & \ar[uuuurr]\ar[rr]_{-2x}  & &   & 4-8e_{8}                                         & \\
 }
}
\caption{Butterfly scheme to obtain a fast computation of the nonlinearity polynomial coefficients, where $(e_1,\ldots,e_8)=(f(\op_1),\ldots,f(\op_8))$.}
\label{figNLPn3}
\end{figure}

We claim the following theorem without proof (see \cite{CGC-cry-art-BellSimSala14}).
\begin{theorem}\label{thmNLPn2n}
  Algorithm \ref{algNLPcoeff} requires
  $O(n2^n)$ integer sums and doublings, in particular circa $n2^{n-1}$ integer sums and circa $n2^{n-1}$ integer doublings, and
  the storage of $O(2^n)$ integers of size less than or equal to $2^n$.
\end{theorem}
%
  %
%

In Table \ref{tabNLtimings} we report the coefficients of growth of the analyzed algorithm and the standard algorithm which uses the fast Walsh transform\footnote{To compute the values in the columns FWT and NLP+FPE we tested $15000$ random \Bf's.}, comparing them with the value $\log_2\big[\frac{(n+1)2^{n+1}}{n2^{n}}\big]$. For each algorithm we compute the average time $t_n$ to compute the nonlinearity of a \Bf with $n$ variables and the average time $t_{n+1}$ to compute the nonlinearity of a \Bf with $n+1$ variables. Then we report in the table the expected theoretical value $\log_2\big(\frac{t_{n+1}}{t_{n}}\big)$.
\begin{table}[h]
\begin{center}
\scalebox{0.8}{
\begin{tabular}{c|*8{c}}

$n$     & 4-5 & 5-6 & 6-7 & 7-8 & 8-9 & 9-10 & 10-11 \\
\hline
$\log_2\big[\frac{(n+1)2^{n+1}}{n2^{n}}\big]$ &
          1.22 & 1.17 & 1.14 & 1.12 & 1.11 & 1.09 & 1.09 \\
FWT     & 0.90 & 0.98 & 1.01 & 1.22 & 0.95 & 1.25 & 1.07 \\
NLP+FPE & 1.02 & 1.09 & 1.13 & 1.07 & 1.17 & 1.07 & 1.11 \\
\end{tabular}
}
\end{center}
\caption{Experimental comparisons of the coefficients of growth of the analyzed algorithm.}
\label{tabNLtimings}
\end{table}
Thanks to theorem \ref{thmNLPn2n} and known facts on fast polynomial evaluation, we obtain:
\begin{theorem}
 Determining the coefficients of the polynomial $\nP_f$ from the truth table of $f$ and then finding $\N(f) = \min\{\nP_f(\bar{a}) \mid \bar{a} \in \{0,1\}^{n+1} \}$ requires a total
 $O(n2^{n})$
 integers operations (sums and doublings).
\end{theorem}
%


  

\section{Acknowledgments}
  \label{secAck}
 The author would like to thank his supervisor, Massimiliano Sala. An extended version of this note, jointly with M.~Sala and I.~Simonetti, can be found at \cite{CGC-cry-art-BellSimSala14}.


\bibliographystyle{alpha}
\bibliography{RefsCGC}




\end{document}